\documentclass{aastex631}
\usepackage{amsmath}
\usepackage{amsbsy}
\accepted{by ApJL}

\shorttitle{Squeezing Drives Southward Magnetic Field Enhancement}
\shortauthors{Wang et al.}

\begin{document}

\title{Squeezing Drives Southward Magnetic Field Enhancement During 10 May 2024 CME-CME Interaction}

\correspondingauthor{Liping Yang, Xushang Feng}
 \email{lpyang@swl.ac.cn, fengx@spaceweather.ac.cn}

\author{Ziwei Wang}
\affiliation{Institute Shenzhen Key Laboratory of Numerical Prediction for Space Storm, School of Aerospace, Harbin Institute of Technology (Shenzhen), Shenzhen 518055, People's Republic of China}
 
\author[0000-0003-4716-2958]{Liping Yang}
\affiliation{School of Geophysics and Information Technology, China University of Geosciences (Beijing), China}

\author[0000-0001-8605-2159]{Xueshang Feng}
\affiliation{SIGMA Weather Group, State Key Laboratory of Solar Activity and Space Weather, National Space Science Center, Chinese Academy of Sciences, Beijing 100190, People's Republic of China}
\affiliation{Institute Shenzhen Key Laboratory of Numerical Prediction for Space Storm, School of Aerospace, Harbin Institute of Technology (Shenzhen), Shenzhen 518055, People's Republic of China}

\author[0000-0001-8179-417X]{Jiansen He}
\affiliation{School of Earth and Space Sciences, Peking University, 100871 Beijing, People's Republic of China}

\author[0000-0002-1369-1758]{Hui Tian}
\affiliation{School of Earth and Space Sciences, Peking University, 100871 Beijing, People's Republic of China}

\author[0000-0002-4935-6679]{Fang Shen}
\affiliation{SIGMA Weather Group, State Key Laboratory of Solar Activity and Space Weather, National Space Science Center, Chinese Academy of Sciences, Beijing 100190, People's Republic of China}
\affiliation{College of Earth and Planetary Sciences, University of Chinese Academy of Sciences, Beijing 100049, People's Republic of China}

\author[0000-0003-4804-5673]{Zhenyong Hou}
\affiliation{School of Earth and Space Sciences, Peking University, 100871 Beijing, People's Republic of China}

\author[0000-0001-5205-1713]{Rui Wang}
\affiliation{State Key Laboratory of Solar Activity and Space Weather, National Space Science Center, Chinese Academy of Sciences, Beijing 100190, People's Republic of China} 

\author{Xiao Zhang}
\affiliation{SIGMA Weather Group, State Key Laboratory of Solar Activity and Space Weather, National Space Science Center, Chinese Academy of Sciences, Beijing 100190, People's Republic of China}
\affiliation{College of Earth and Planetary Sciences, University of Chinese Academy of Sciences, Beijing 100049, People's Republic of China} 

\author[0000-0001-9427-7366]{Ming Xiong}
\affiliation{SIGMA Weather Group, State Key Laboratory of Solar Activity and Space Weather, National Space Science Center, Chinese Academy of Sciences, Beijing 100190, People's Republic of China}
\affiliation{College of Earth and Planetary Sciences, University of Chinese Academy of Sciences, Beijing 100049, People's Republic of China}

\begin{abstract}

The southward magnetic field carried by coronal mass ejections (CMEs) is a key driver of space weather events, yet how interacting CMEs amplify this field remains debated. The 10 May 2024 event, which produced the strongest geomagnetic storm since 2003, offers an ideal testbed for this question. Combining multi-spacecraft observations with high-resolution controlled numerical experiments, we track the CMEs continuously from their origin near the solar surface to beyond Earth, and find that inherent squeezing during CME-CME interactions generates the sustained enhancement of the southward field, exceeding 60~nT in magnitude near Earth. First driven by the dynamic pressure of the trailing CME and later maintained by magnetic and thermal pressures at the CME-CME interface, squeezing transforms the preceding CME from a bulky cloud into a flattened ``pancake'' structure, prevents expansion-induced field decay, and maintains an amplified southward field over long distances. The strength of this enhancement scales with the intensity of the squeezing. Our findings establish a direct link between CME-CME interaction and strong southward fields, laying a physics-based foundation for understanding and forecasting the geoeffectiveness of interacting CMEs.

\end{abstract}


\section{Introduction} \label{sec:intro}
Coronal mass ejections (CMEs) are the most violent eruptive phenomena in the solar atmosphere, releasing magnetized plasma blobs with masses of \(10^{15}\)--\(10^{16}\)~g and energies of \(10^{23}\)--\(10^{25}\)~J into the interplanetary space \citep{Chen_2011, Luhmann2020}.  
The southward magnetic field carried by Earth-directed CMEs is widely recognized as a primary driver of severe space weather events \citep{Tsurutani1988, Dai2024}, which can disrupt communications, navigation, power grids, and space activities, incurring substantial socioeconomic costs \citep{Borovsky2006}. Elucidating how this southward magnetic field forms, amplifies, and evolves is therefore a central goal in heliospheric physics and a prerequisite for improving forecasts of space weather events. 
 
When two or more CMEs erupt in succession, they can interact in interplanetary space, sometimes enabling a moderately geoeffective CME to trigger an extreme geomagnetic storm after interaction \citep{Lugaz2017, Zhang2021}. Indeed, several iconic superstorms in history -- the 1859 Carrington event, the 1989 Quebec event, the 2003 Halloween event, and the May 2024 Gannon event -- have all been closely associated with the interaction among multiple CMEs. Interacting CMEs are often observed to carry a strong southward magnetic field, making them more likely to induce intense and long-lived geomagnetic activity than individual CMEs \citep{Burlaga1987, Farrugia2004, Shen2012, Liu2014, Gopalswamy2025}.  

Shock compression is widely considered to be the mechanism responsible for the strong southward magnetic field in CME-CME interaction. When a shock driven by a trailing fast CME catches up and propagates into a preceding slow CME, a shock-CME structure forms \citep{Ivanov1982, Wang2003a, Lugaz2015ab}. Within such a structure, the shock compresses the magnetic field inside the CME and thereby amplifies its southward field \citep{Behannon1991, Liu2012,  Lugaz2015, Xu2019, Chi2024}. Meanwhile, numerical simulations have been conducted to understand and quantify the impact of a shock on a preceding CME, and have shown that the intensity of the southward magnetic field can be amplified by a factor of two or more \citep{Lugaz2007, xiong2009, Scolini2020, Koehn2022, Wijsen2025}. However, shock compression alone seems insufficient for the prolonged nature of the strong southward field in many extreme storms.

Apart from shock compression, other processes have been proposed to operate during CME-CME interaction and may induce significant variations in CMEs \citep{Temmer2012, Manchester2017, Vourlidas2019}. 
For instance, when two CMEs are sufficiently close, they may squeeze and deform each other, and such deformation may be a cause for the increase in the internal magnetic field strength of CMEs \citep{Xue2005, Song2026}. In addition, magnetic reconnection between CMEs has also been hypothesized, potentially leading to the merging of two CMEs---a process known as ``CME cannibalism'' \citep{Gopalswamy2001, Xiong2007, Koehn2022, Biswas2025}, which might result in an isolated, long-lasting magnetic-cloud-like structure that could sustain prolonged geomagnetic activity \citep{Dasso2009, Lugaz2014, Chen2024}. 

However, it remains largely unclear whether and how shocking, squeezing, and merging actually contribute to the southward magnetic field enhancement during CME-CME interactions. This uncertainty mainly arises from the fact that these processes often operate simultaneously and are difficult to isolate in limited observations. Disentangling their respective contributions is therefore essential for identifying the underlying mechanism responsible for the enhanced southward field, a crucial step toward understanding and predicting the geoeffectiveness of CMEs.

The May 2024 CME interaction event provides an ideal opportunity to resolve how these coexisting processes collectively produce the enhanced southward field. 
On 8-9 May 2024, solar active region (AR) 13664 produced a series of X-class flares accompanied by multiple Earth-directed CMEs, which successively chased each other in interplanetary space \citep{WangR2024, Li2024, Liu2024, Hayakawa2025, Khuntia2025}. Studies show that the pile-up and interaction of these CMEs eventually triggered the strongest geomagnetic storm since 2003 ($Kp = 9$, $Dst \approx -412$ nT)  \citep{Kwak2024, Fu2025, Chen2025, Sabri2025, Zhang2026ab, Ren2026}. However, systematic analyses are still lacking on how the interaction produces the observed complex ejecta structure with its remarkably strong southward magnetic field.

To address this missing link, this paper combines multi-spacecraft observational data with high-resolution numerical simulations to investigate how the southward magnetic field is modulated during  the interaction event on 10 May 2024. It is found that squeezing occurring in the interaction of CMEs both generates and sustains a strong southward magnetic field over long distances. This mechanism explains why interacting CMEs during this event can maintain such a strong southward field when no internal shock is detected. The paper is structured as follows: Section 2 describes the modeling methodology; Section 3 presents the analysis of the results; and Section 4 summarizes and discusses the findings.

\section{Modeling setup}

In this study, we employ three-dimensional Solar-Interplanetary (SIP) adaptive mesh refinement
(AMR) space-time conservation element and solution element (CESE) solar wind model
(SIP-AMR-CESE Solar Wind Model) developed by \cite{Feng2012} and \cite{Yang2012} to simulate the propagation and evolution of the background solar wind and CMEs from the corona to interplanetary space. The computational domain covers $1~Rs \leq r\leq 250~Rs$, $0^\circ \leq \theta \leq 180^\circ$, and $0^\circ \leq \phi \leq 360^\circ$. Here, $Rs$ is solar radius, $\theta$ is  colatitude angle, and $\phi$ is  longitude angle.  

We initialize the ambient solar wind by adopting the Parker hydrodynamic solar wind solution and a potential field derived from the integral Global Oscillation Network Group (GONG) synoptic magnetogram for Carrington Rotation 2284. The temperature and number density at the solar surface are set to  $1.45\times10^6$ K and $2\times10^8$ cm$^{-3}$, respectively. For the boundary conditions, the variables at the inner boundary $1\ Rs$ are fixed for simplicity, while at the outer boundary, where the flow becomes supersonic and super-Alfv\'enic, an extrapolation scheme is applied. The governing MHD equations are integrated in time until the solar wind flow and magnetic field reach a quasi-steady state. This quasi-steady-state solar wind then serves as the background for the subsequent propagation of the CMEs.

To initiate a CME, we superpose a spheromak on the quasi-steady-state solar wind, owing to its simplicity and widespread adoption \citep{Scolini2020, Yang2021, Yang2025}.
For the simulation of specific events, the parameters of the spheromak model are determined by the CME parameters derived from the GCS reconstruction \citep{Yang2023, Ma2024, Zhang2026}. Practically, the propagation speed of the CME defines the initial speed of the spheromak. Since the CME propagation directions are generally consistent with their source locations on the Sun, the source location is used as the center position of the spheromak, which is placed at 
2 $R_s$. 
The CME center-axis locations derived from the GCS reconstructions are consistent with their surface source locations to within a few degrees (e.g., \cite{Liu2024}). Here, we adopt the source location as the center position of the spheromak as this provides a single, self-consistent anchor for all three CMEs that is directly tied to the eruption site in AR 13664.
The aspect ratio prescribes the radius of the spheromak. The polar axis direction of the spheromak is constrained by the tilt angle, and the magnetic field strength of the spheromak is iteratively adjusted to match both white-light observations and in-situ measurements. Based on these remote-sensing and in-situ observational constraints, the three CMEs are injected sequentially into the background solar wind, with time intervals consistent with those at which each CME first appears in the SOHO/LASCO-C2 FOV. 
 
To investigate the influence of CME-CME interaction on the enhancement of $B_z$, we design three comparative cases.
In Case A, which mimics the homologous nature of CME1, CME2, and CME3, the initial magnetic helicities of the three CMEs are set to be the same, so that all carry a southward magnetic field component. Because the magnetic field directions at the tail of CME1 are opposite to those at the head of CME2, the interaction between CME1 and CME2 is dominated by merging. We refer to Case A as the ``Merged CME1 and CME2'' case. Case B is based on Case A, but the initial magnetic helicity of CME2 is reversed (opposite to that of CME1). As a result, the magnetic field directions at the tail of CME1 are the same as those at the head of CME2, and the interaction between CME1 and CME2 is expected to be dominated by squeezing. Case B is therefore called the ``Squeezed CME1 and CME2'' case. Case C (``Only CME1'') injects only CME1 into the background solar wind, serving as a blank control without any subsequent CMEs or CME-CME interaction.
These cases are designed as controlled experiments to isolate the physical contribution of squeezing rather than reproduce all magnetic details of the event as the spheromak model is used here to capture macroscopic magnetic configuration of CMEs.

To efficiently handle the large spatial scales in heliospheric simulations while maintaining high resolution in regions of interest, we resolve the computational volume using a block‑structured grid with AMR. This refinement is designed to automatically capture CMEs and their associated key structures, such as shocks, sheaths, and current sheets, at the high resolution.
The refinement criteria are based on the passive tracer $\rho_c$ and the gradients and curls of velocity and magnetic field. Specifically, regions with 
$\rho_c>0$ track the CME material and the sheath, the divergence of velocity captures shocks, and the normalized curl of the magnetic field identifies regions of strong electric current. When a criterion is satisfied, the corresponding block is refined; after the CME passes, the block is coarsened back to the base resolution. With these settings, six levels of grid refinement are achieved. This strategy allows the CMEs to be traced from the solar corona to Earth with manageable computational cost.

\section{Results}

\subsection{EUV and White-light Signatures}\label{subsec1}

\begin{figure}[htbp]
\centering
\includegraphics[width=1\textwidth]{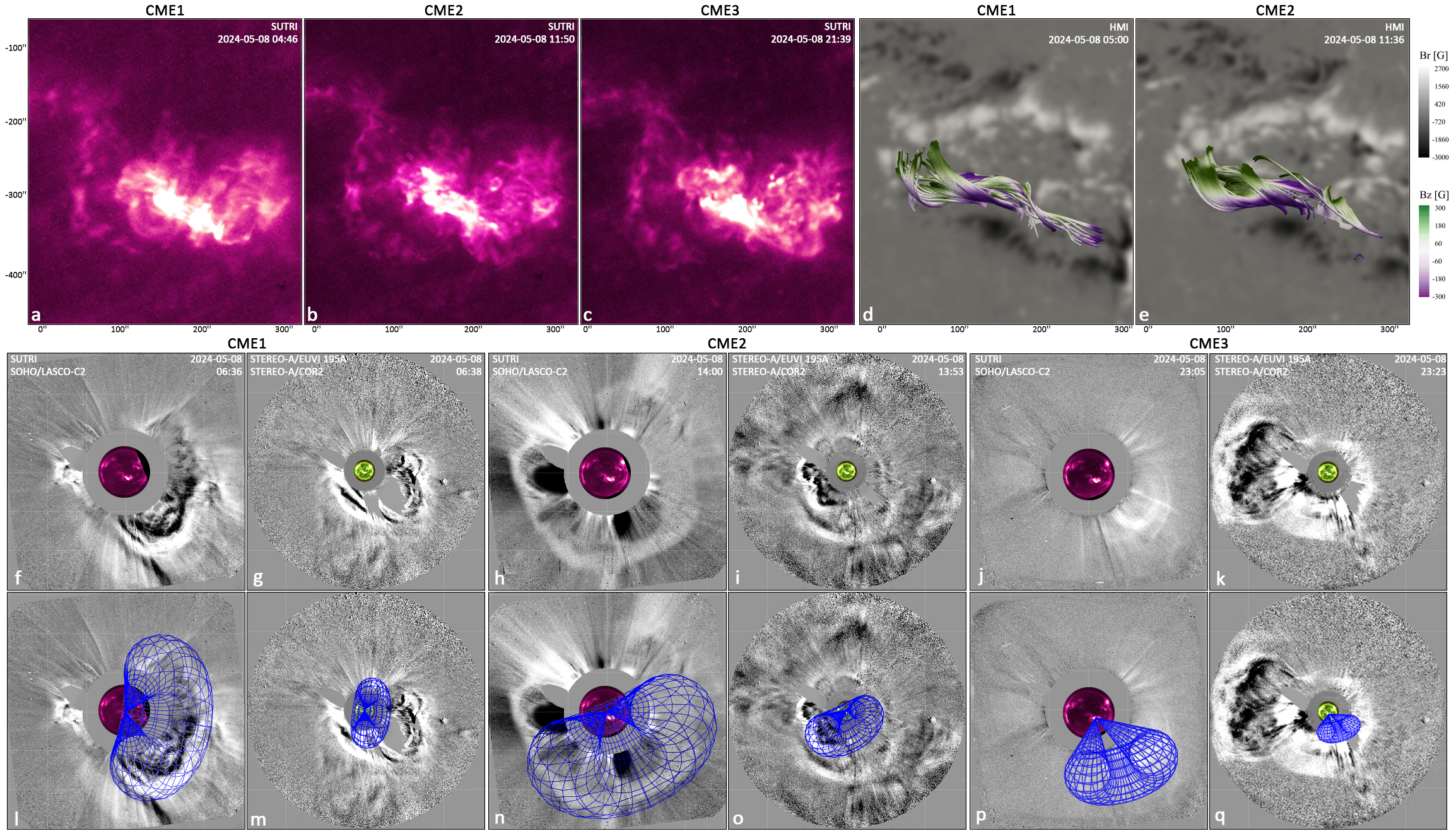}
\caption{\textbf{Eruption and propagation of CME1, CME2, and CME3 in the solar corona.} Panels a-c show the EUV imagery from SUTRI for each event on 8 May 2024. Panels d and e present field lines overlaid on the SHARP (Space‑weather HMI Active Region Patch) magnetogram. The field lines are derived from NLFFF extrapolations, revealing flux-rope structures. The color of the field lines indicates the $z-$component of the magnetic field $B_z$. Panels f-k display running-difference images of  CME1, CME2, and CME3 from SOHO/LASCO-C2 and STEREO-A/COR2, together with EUV observations from SUTRI and STEREO-A/EUVI 195 \,\AA. Panels l-q illustrate the fitted GCS model overlaid as blue wireframes based on the observations in panels f-k. An animation of this figure is available. It shows the EUV observations from SDO/AIA 94 \AA\ depicting the eruption of the three successive CMEs from AR 13664 on 2024 May 8.}\label{fig1}
\end{figure}

Combining the EUV observations from SUTRI (Solar Upper Transition Region Imager) and SDO/AIA (Solar Dynamics Observatory/Atmospheric Imaging Assembly) 94 \,\AA\ (the animation in Fig. 1) with the white-light observations from SOHO/LASCO (Solar and Heliospheric Observatory/Large Angle and Spectrometric Coronagraph)-C2 and STEREO-A (Solar Terrestrial Relations Observatory - Ahead)/COR2, four eruptions associated with four CMEs occurred on 8 May 2024. The first, second, and fourth eruptions originated from solar active region (AR) 13664, and were associated with major soft X-ray flares of classes X1.0, M8.3, and X1.0/M9.8, respectively, whereas the third eruption originated from AR13667 and was related to a filament eruption \citep{WangR2024, Khuntia2025, Weiler2025}. The corresponding first, second, and fourth CMEs were observed as full-halo CMEs, making them the most likely candidates to have impacted Earth, while the third CME propagated in a direction toward the west-north of Earth and was probably not an Earth-directed CME. Given that the first, second, and fourth CMEs erupted successively from the same active region and propagated in nearly the same direction, they likely interacted during their journey to Earth. Here, we name them CME1, CME2, and CME3 (Fig.~\ref{fig1}) and focus on their interaction processes as well as the resulting geoeffectiveness, using both observations and simulations.
 
Each of CME1, CME2, and CME3 was associated with impulsive loop eruptions observed by SUTRI (Ne VII 46.5~nm line) and SDO/AIA 94\,\AA\ (Fig.~\ref{fig1}a-c), which have peak formation temperatures of about 0.5~MK and 7~MK, respectively. Focusing on the magnetic field configurations associated with the impulsive loop eruptions, the NLFFF (non-linear force-free field) extrapolation reveals strongly sheared field lines that form flux-rope structures and run along the magnetic polarity inversion line (Fig.~\ref{fig1}d--e). Consistent with observations of flare ribbons and coronal dimmings \citep{Jarolim2024, Weiler2025, Razquin2026}, the flux ropes have a low inclination with respect to the solar equator. 
Dominated by the southward  $B_z$ (the $z-$component of the magnetic field), these flux systems participated in the eruptions, and one would expect that the resulting CMEs carried a strong southward $B_z$ during their early stage. 

It is noted that the orientation of the flux rope for CME1 is similar to that for CME2, indicating that CME1 and CME2 are homologous CMEs. As no vector magnetic field observations are available for CME3 from SDO/HMI (Helioseismic and Magnetic Imager), we cannot determine its flux-rope structure. 
However, the impulsive loop eruptions observed by SUTRI and SDO/AIA, together with the shared PIL location and consistent magnetic field configuration \citep{WangR2024}, strongly suggest that CME1, CME2, and CME3 are homologous eruptions.
These three CMEs are considered homologous events, as they erupted successively from the same active region (AR 13664) within a short time interval \citep{Lugaz2017}. Their tilt angles vary for the events, introducing asymmetry into the interaction. Our simulation captures this asymmetry by initializing the spheromak tilt of each CME separately from its individual GCS reconstructions, rather than assuming identical geometries.

By combining coronagraph observations from SOHO/LASCO-C2 and STEREO-A/COR2 with EUV observations from SUTRI and STEREO-A (Fig.~\ref{fig1}f--k), the three-dimensional morphology of the CMEs is reconstructed using the Graduated Cylindrical Shell (GCS) technique \citep{Thernisien2006} by assuming a rope-like structure for each of CMEs with both ends anchored in AR13664. Since STEREO-A was located about 12$^\circ$  west of the Sun-Earth line,  CME1, CME2, and CME3 are visible as halo CMEs and can be separately identified from both the SOHO and STEREO-A viewpoints, which facilitates the reconstruction. The derived orientation parameters of CME1, CME2, and CME3 show no substantial change within the observed field of view (FOV), evidencing again that CME1, CME2, and CME3 are homologous CMEs. The derived parameters of CME1, CME2, and CME3 are also used to constrain the initialization of the CMEs in our simulations.

\begin{figure}[htbp]
    \centering
    \includegraphics[width=\textwidth]{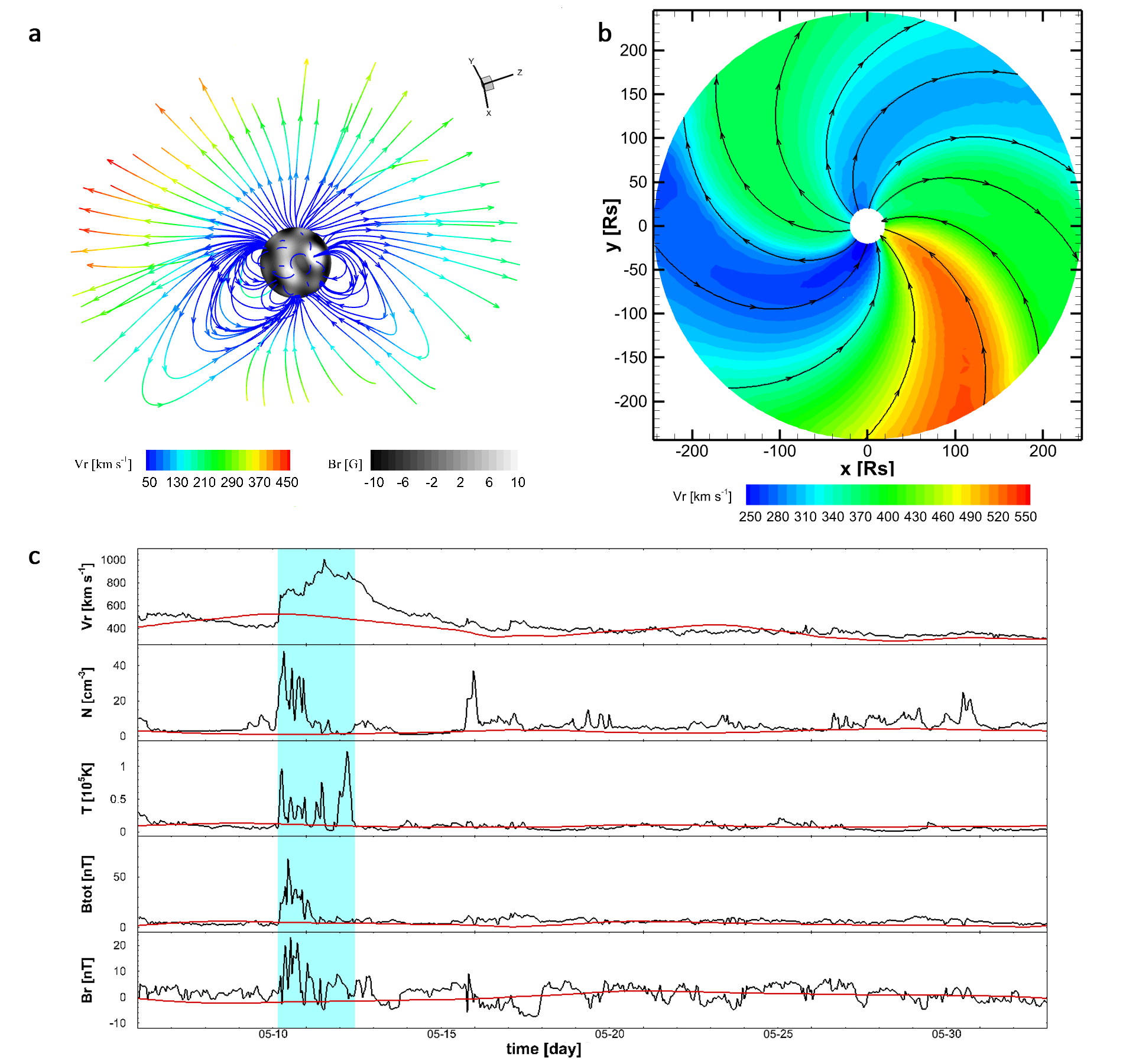} 
    \caption{\textbf{Background solar wind simulation results.} Panel a shows the distribution of three-dimensional magnetic field lines from 1 to 6 \(R_s\). Panel b shows the velocity distribution in the ecliptic plane within 20--215 \(R_s\). Panel c presents the time series comparison between the simulated results (radial velocity \(V_r\), number density \(N\), temperature \(T\), total magnetic field strength \(B_{tot}\), and radial magnetic field component \(B_r\)) and the OMNI data. The simulated results are shown as red solid lines, and the OMNI data as black solid lines. The shaded regions indicate the intervals of CME disturbances at 1 AU.}\label{fig2}
\end{figure}

Fig.~\ref{fig2} validates the quality of the background solar wind simulation. The large-scale solar wind structures are reproduced (Fig.~\ref{fig2}a and b). The time-series comparison with OMNI data (Fig.~\ref{fig2}c) shows that the simulated radial velocity, number density, temperature, and magnetic field parameters are basically in agreement with the in-situ observations during the quiet periods surrounding the CME disturbance. These comparisons confirm that the steady-state solar wind provides a reliable background for the CME simulations.

\begin{figure}[htbp]
    \centering
    \includegraphics[width=1\textwidth]{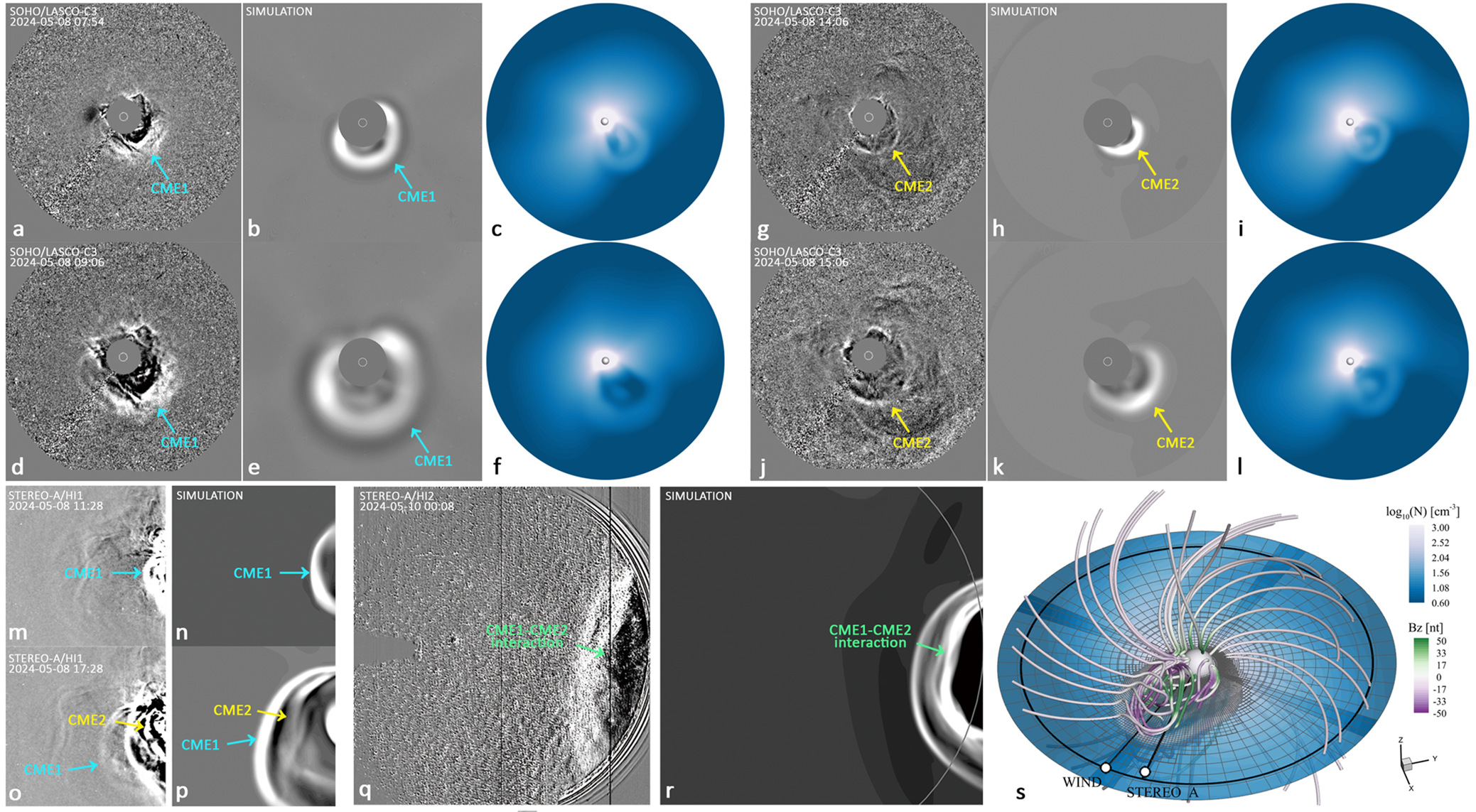}
    \caption{\textbf{Observed and simulated white-light morphologies of CME1 and CME2 in the inner heliosphere.} Panels a, d, g, and j show running-difference images from SOHO/LASCO-C3 coronagraph observations at different times. Panels b, e, h, and k are the corresponding synthetic ones from the simulation. Panels c, f, i, and l display the density distribution from 1 $R_s$ to 30 $R_s$. Panels m-p present running-difference images from STEREO-A/HI1 observations alongside their synthetic counterparts at different times. Panels q and r show running-difference images from STEREO-A/HI2 observations and the corresponding synthetic images. Panel s displays a top view of the simulated CMEs, including the distribution of their magnetic field lines, with colors indicating the  magnetic field component $B_z$. The blue plane in panel s denotes the density distribution on the ecliptic plane from 20 $R_s$ to 250 $R_s$, where the gray quadrilaterals represent the computation blocks, with each block subdivided into $6\times6\times6$ grid cells. An animation of this figure is available. The animation runs in serial and shows the continuous propagation of the CMEs across different fields of view: first, the outer corona from SOHO/LASCO-C3; second, the interplanetary space from STEREO-A/HI1; and finally, the heliospheric evolution from STEREO-A/HI2 from May 9 to 10.}\label{fig3}
\end{figure}

Fig.~\ref{fig3} and the corresponding animation show that in the FOV of SOHO/LASCO-C3, CME1 and CME2 are still clearly separated and can be easily identified as halo CMEs. On 8 May 2024 at 07:54 UT(Universal Time), CME1 shows an enhanced brightness in the south-westerly direction (Fig.~\ref{fig3}a), and by 09:06 UT it has fully entered the SOHO/LASCO-C3 FOV. The front of CME2 appears weaker than that of CME1 but remains clearly visible, also propagating south-westward (Fig.~\ref{fig3}g and j). Constrained by observations from SOHO/LASCO-C2 and STEREO-A/COR2, the simulations successfully reproduce the morphologies of the observed CMEs in SOHO/LASCO-C3 images (Fig.~\ref{fig3}b, e, h, and k), with the density showing a clear enhancement at the leading fronts of both CME1 and CME2 (Fig.~\ref{fig3}c, f, i, and l).
 
Beyond the SOHO/LASCO-C3 FOV, the white-light signatures of the CMEs are also captured by STEREO-A/HI1 (Fig.~\ref{fig3}m and o) and STEREO-A/HI2 (Fig.~\ref{fig3}q).
Fig.~\ref{fig3}m, o and the corresponding the animation in Fig. 3 show that the distinct fronts of CME1 and CME2 are clearly identified within the STEREO-A/HI1 FOV, and CME2 has not yet caught up with CME1. However, the STEREO-A/HI2 observations show that the fronts of CME1 and CME2 appear to merge into a large, bright front, meaning that the two CMEs pile up and influence each other (Fig.~\ref{fig3}q and the animation in Fig. 3). Meanwhile, the shapes of CMEs seem to transform from a bulky cloud (within the SOHO/LASCO-C3 and STEREO-A/HI1 FOV) into a flattened ``pancake'' structure. The corresponding synthetic images from the simulation generally agree with these observations. The position, angular width, and propagation direction of the simulated CMEs roughly consistent with those observed by STEREO-A/HI1 and HI2, providing a solid foundation for the subsequent analysis of the CME interaction process.

\subsection{Interplanetary Interaction}\label{subsec2}

\begin{figure}[htbp]
    \centering
    \includegraphics[width=0.6\textwidth]{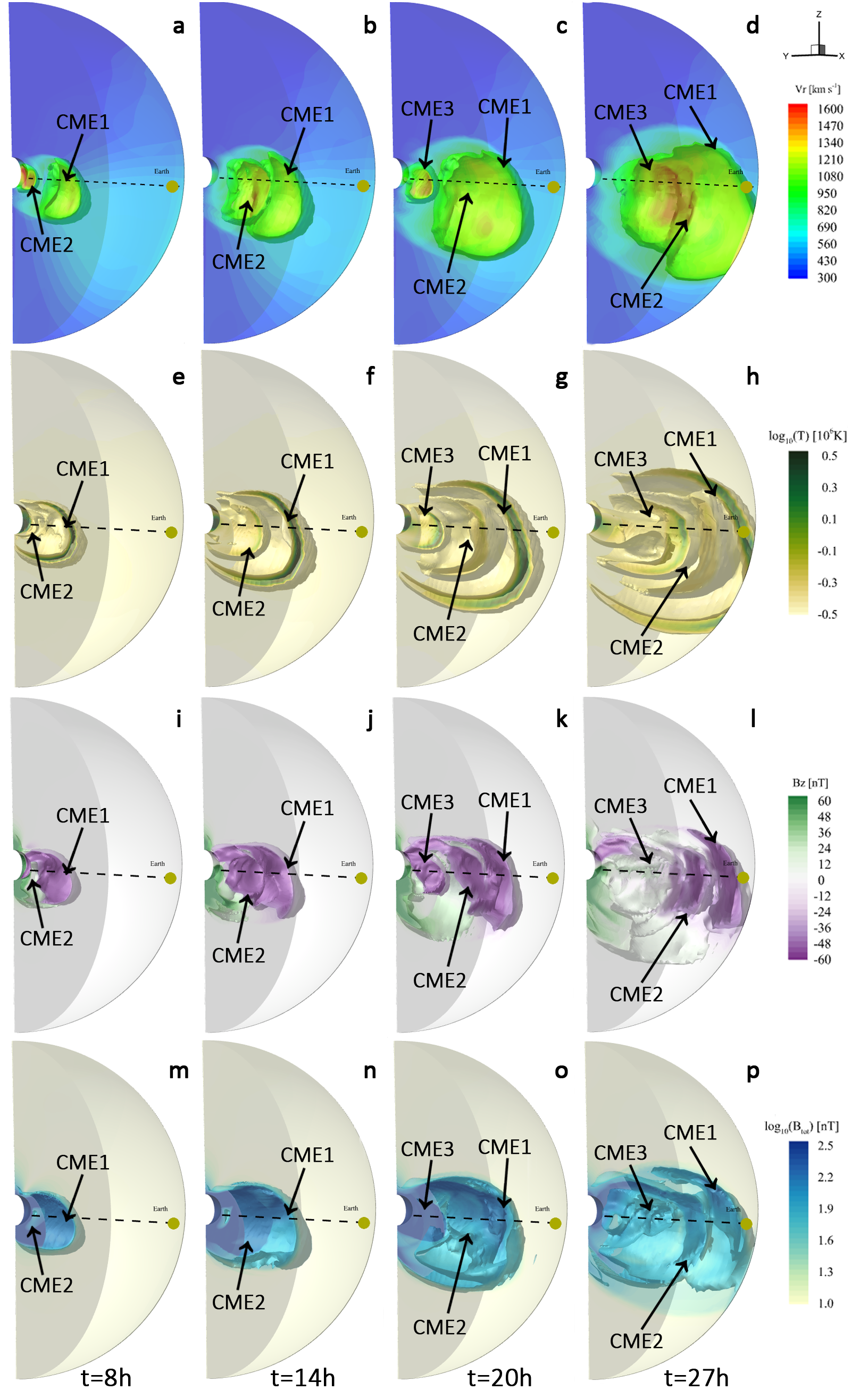}
    \caption{\textbf{Propagation and interaction of the three CMEs from 20 $R_s$ to 250 $R_s$.} Panels a-d show the radial velocity ($V_r$), panels e-h present the temperature ($T$), panels i-l display the southward magnetic field component ($B_z$), and panels m-p show the total magnetic field strength ($B_{\rm tot}$) at different times for the case labeled ``Merged CME1 and CME2''. An animation of this figure is available. The animation runs in parallel (mosaicked into three rows) to compare three different cases simultaneously: the ``Merged CME 1 and CME 2'' scenario, the ``Squeezed CME 1 and CME 2'' scenario, and the ``Only CME 1'' scenario. It displays the evolution of the radial velocity (\(V_r\)), temperature (\(T\)), magnetic field component (\(B_z\)), and total magnetic field strength (\(B_{\rm tot}\)) in the interplanetary space from 20 \(R_s\) to 250 \(R_s\).}\label{fig4}
\end{figure}

Fig.~\ref{fig4} and the top row of the animation in Fig. 4 show the evolution of the three CMEs as they propagate and interact in interplanetary space. 
At $t=8$ h, CME1 and CME2 are well separated and propagate independently. With a speed of about 1000 km s$^{-1}$ (Fig.~\ref{fig4}a), CME1 drives a shock at its leading edge (Fig.~\ref{fig4}e). Meanwhile, CME1 is dominated by a southward $B_z$ along the Sun-Earth line (Fig.~\ref{fig4}i), which originates from the eruption stage (Fig.~\ref{fig1}). CME2 moves faster than CME1 and therefore chases it, as they are homologous eruptions propagating in similar directions (Fig.~\ref{fig4}a, Figs.~\ref{fig1} and~\ref{fig3}).

Approximately six hours later, CME2 comes into contact with CME1. At this stage, although CME2 can still be clearly distinguished from CME1, the interaction between the two CMEs begins (Fig.~\ref{fig4}b, f, j, n). By $t=20$ h, CME1 and CME2 are interacting intensely, and as a result, they travel at almost the same speed (Fig.~\ref{fig4}c). 
Joint analysis of temperature T and $B_z$ (Fig.~\ref{fig4}g, k, and Fig.~\ref{fig4} and the top row of the animation in Fig. 4) identifies a hot interface between CME1 and CME2 that separates the two structures. In the squeezed CME interaction case (the middle row of the animation in Fig. 4), $B_z$ polarity provides a robust tracer to distinguish CME1 and CME2 despite the persistent hot contact layer.

Compared to the ``Only CME1'' case (the bottom row of the animation in Fig. 4), CME1 in both the ``Merged'' and ``Squeezed'' scenarios is significantly squeezed by CME2, with its radial extent compressed to a very small range and a flattened ``pancake'' structure forming (Fig.~\ref{fig4} and the top and middle row of the animation in Fig. 4).  Consequently, the 
$B_z$ inside CME1 does not decrease due to expansion; instead, it remains strong. Over the next few hours, the interaction continuously compresses CME1, and a strong southward $B_z$ (exceeding $-60$~nT) is maintained in its interior (the top and middle row of the animation in Fig. 4). At  $t=27$ h, the front of CME1 reaches Earth (Fig.~\ref{fig4}d, h, i, and p). Meanwhile, CME3 begins to interact with CME2.

\begin{figure}[htbp]
    \centering
    \includegraphics[width=0.6\textwidth]{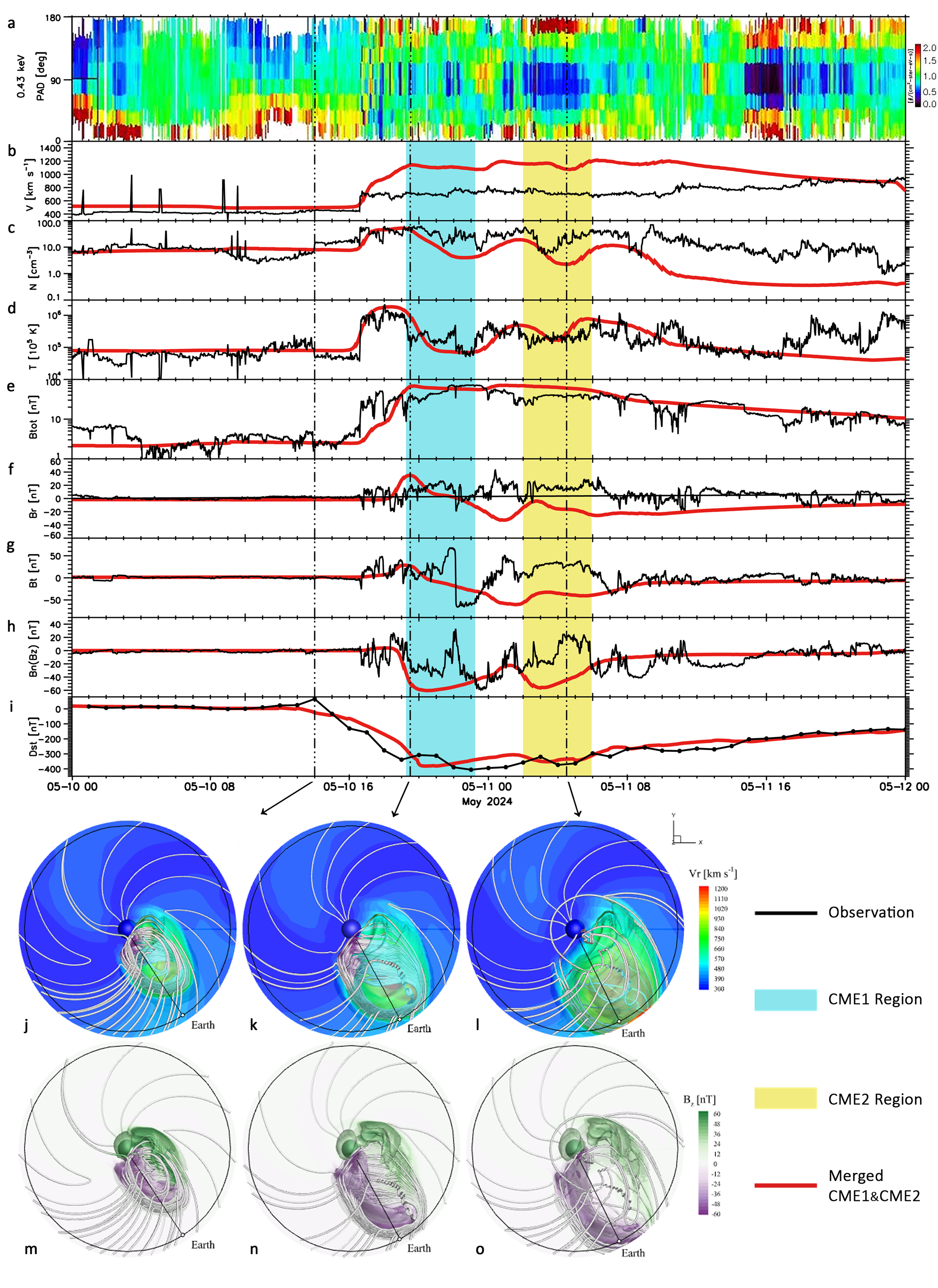}
    \caption{\textbf{The complex structure observed at WIND and sampled from the simulation.} Panel a shows the electron pitch-angle distribution. Panels b-i present the time series of speed ($V$), number density ($N$), temperature ($T$), total magnetic field strength ($B_{\rm tot}$), the magnetic field components ($B_r$, $B_t$, $B_n\ (B_z)$), and the Dst index. The black solid line represents the observations, and the red solid lines correspond to the simulation case of ``Merged CME1 and CME2''. The shaded regions indicate the possible intervals of the observed CME1 and CME2. The simulation data are shifted by about 11 hours to match the observations.  Panels j-l and m-o display the distributions of the velocity $V$ and the magnetic field component $B_z$ on the ecliptic plane at three representative times marked by the dash-dotted lines. The gray lines therein denote mangetic field lines. An animation of this figure is available. The animation runs in parallel (mosaicked into three rows) showing the evolution of the CMEs on the ecliptic plane from 20 \(R_s\) to 250 \(R_s\) for the three scenarios: ``Merged CME 1 and CME 2'', ``Squeezed CME 1 and CME 2'', and ``Only CME 1''. Displayed quantities include the radial velocity (\(V_r\)), temperature (\(T\)), magnetic field component (\(B_z\)), and total magnetic field strength (\(B_{\rm tot}\)).}\label{fig5}
\end{figure}

The CME-CME interaction results in the complex structure formed at 1~AU, which is revealed by the WIND measurements and corresponding simulation results (Fig.~\ref{fig5}).
Here we show only the ``Merged CME1 and CME2'' case, which is designed to mimic the actual event, while the ``Squeezed CME1 and CME2'' case is used as a controlled numerical experiment to highlight the squeezing effects. In contrast to the quiet solar wind, the complex structure begins with an interplanetary shock, marked by a significant increase in solar wind speed ($V$), number density (N), temperature (T), and magnetic field strength ($B_\textrm{tot}$). Behind the shock lies a sheath region filled with high-speed, high-density, high-temperature plasma and a strong magnetic field (Fig.~\ref{fig5}b-e). 

The shaded regions in Fig.~\ref{fig5} show the likely intervals corresponding to the cores of CME1 and CME2, which are identified
according to bidirectional suprathermal electron pitch-angle distributions (PADs), low $N$ and $T$, a strong and rotated magnetic field, and further supported by the simulated result showing the high-temperature interface (Fig.~\ref{fig4}g). The region between the cores of CME1 and CME2 is recognized as the high-temperature interface, which lacks bidirectional PADs (Fig.~\ref{fig5}a and d). Fig.~\ref{fig5}b shows that the observed speed difference between CME1 and CME2 is very small, likely a consequence of their interaction, as revealed by the simulations (Fig.~\ref{fig5}j--l and Fig.~\ref{fig4}a--d).

The interaction simulations successfully reproduce large southward $B_z$ near Earth. Its peak value is about $60$~nT for the ``Merged CME1 and CME2'' case (red line in Fig.~\ref{fig5}h) and for the ``Squeezed CME1 and CME2'' case (the middle row of the animation in Fig. 4 and 5), while it is only about 30~nT for the ``only CME1'' case (the bottom row of the animation in Fig. 4 and 5). These results indicate that the CME-CME interaction significantly  amplifies the southward field, causing a minimum Dst of about -400 nT (Fig.~\ref{fig5}i). From the propagation of the CMEs in the ecliptic plane (Fig.~\ref{fig5}m-o and the animation in Fig. 5), the complex structure encounters a head-on collision with Earth, accounting for the very strong southward $B_z$ near Earth. 

\subsection{Southward $B_z$ Enhancement}\label{subsec3}

\begin{figure}[htbp]
    \centering
    \includegraphics[width=0.9\textwidth]{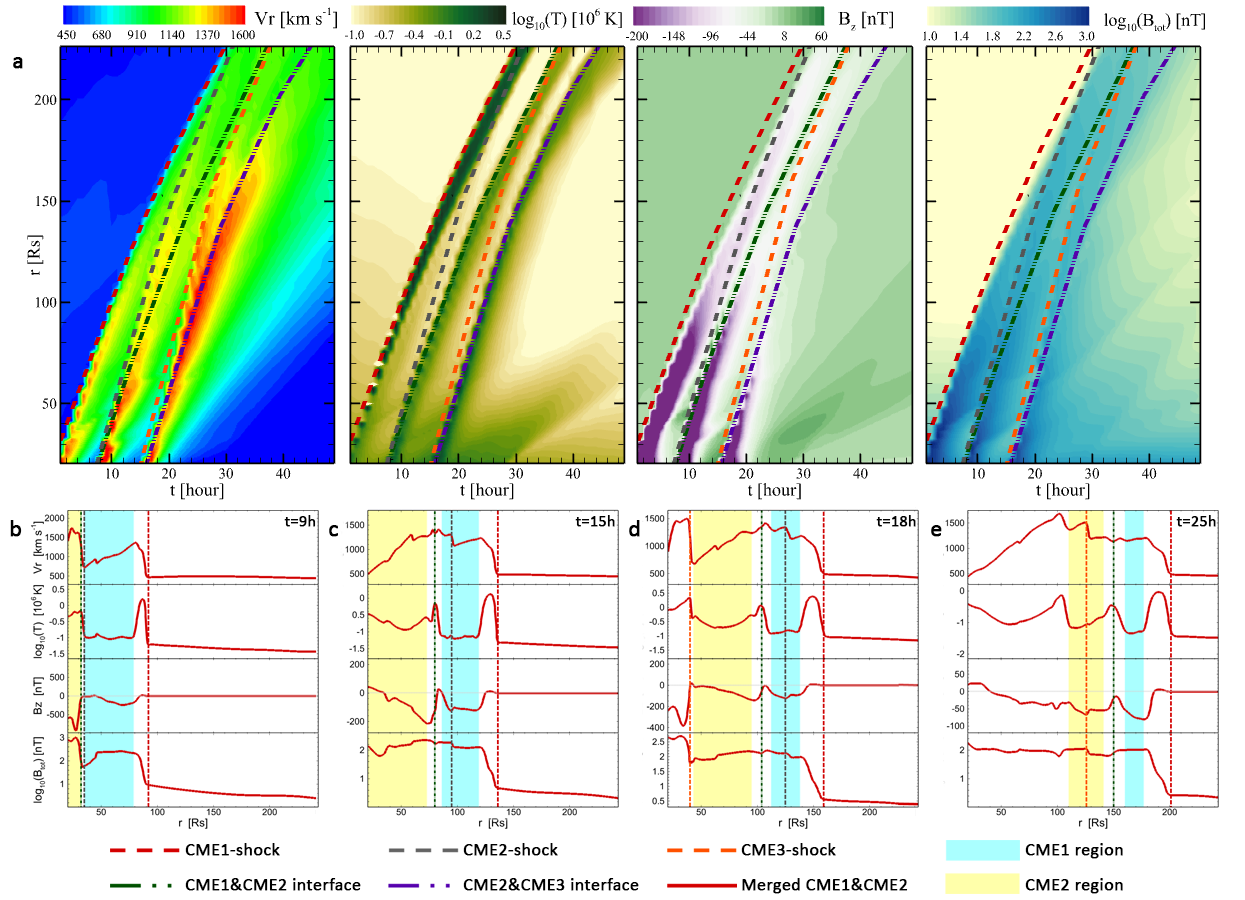}
    \caption{\textbf{Propagation of the three CMEs along the Sun-Earth line.} Panels a-b show the time-space slices of the radial velocity ($V_r$), the temperature ($T$), the magnetic field component $B_z$, and the total magnetic field strength ($B_{\rm tot}$) for the simulation case:  ``Merged CME1 and CME2''. The dashed lines indicate the shock positions, while the dash-dotted lines mark the interfaces between the CMEs. Panels c-f present the radial variations of the same four quantities at four representative times. The red and blue solid lines correspond to the two cases, and  the shaded regions indicate the possible intervals of CME1 and CME2.}\label{fig6}
\end{figure}
 
To characterize the details of the interaction between CME1 and CME2 and the resulted $B_z$ enhancement in CME1, we take one-dimensional (1D) cuts along the Sun-Earth line at various times during our simulations. Note that the Sun-Earth line approximately crosses the center of the CMEs and thus aligns with the propagation direction of the CME fronts (Fig.~\ref{fig5}). For the two cases, ``Merged CME1 and CME2'' and ``Squeezed CME1 and CME2'', we identify the locations of the shocks driven by CME1, CME2, and CME3 by searching for abrupt increases in speed (dashed lines in Fig.~\ref{fig6}); the interfaces between CME1 and CME2 and between CME2 and CME3 by regions of high temperature (dash-dotted lines in Fig.~\ref{fig6}); and the intervals of CME1 and CME2 by low-temperature regions corresponding to ejecta (shaded regions in Fig.~\ref{fig6}). The features of the identified  CMEs and the identified interfaces between the CMEs are consistent with the measurements in Fig.~\ref{fig5}.
 
Fig.~\ref{fig6} shows the shock-ejecta interaction and the induced magnetic enhancement of CME1.
As the shock driven by CME2 (gray dashed lines) propagates into CME1, the magnetic field of CME1 is compressed, forming a dip in the southward 
$B_z$ (Fig.~\ref{fig6}c-d). Because the shock in the ``Squeezed CME1 and CME2'' case is stronger (larger speed jump) than that in the ``Merged CME1 and CME2'' case, the dip in the squeezed case is more pronounced. At about $t=22$ h, the shock reaches near the front of CME1 and begins interacting with the dense sheath plasma (Fig.~\ref{fig6}a).  
 
Fig.~\ref{fig6} also displays the variations of CME1 after the shock  passes through CME1.
At this stage, the radial size of CME1 continuously decreases (Fig.~\ref{fig6}a, e-f), consistent with sustained squeezing by the trailing CME2 or by the interface between CME1 and CME2. This confinement inhibits the expansion of CME1 and slows the decay of 
$B_z$ with distance (Fig.~\ref{fig6}a).
Note that after $t=18$h, the magnetic field in CME1 becomes even stronger than that in CME2 (Fig.~\ref{fig6}e-f), consistent with in-situ measurements (Fig.~\ref{fig5}e). Furthermore, the short radial extent of CME1 resulting from sustained squeezing is consistent with the WIND measurements in Fig.~\ref{fig5}.
Due to being squeezed by CME3, the radial width of CME2 decreases evidently from $t=18$h to $t=25$h (Fig.~\ref{fig6}e-f).
  
\begin{figure}[htbp]
    \centering
    \includegraphics[width=0.95\textwidth]{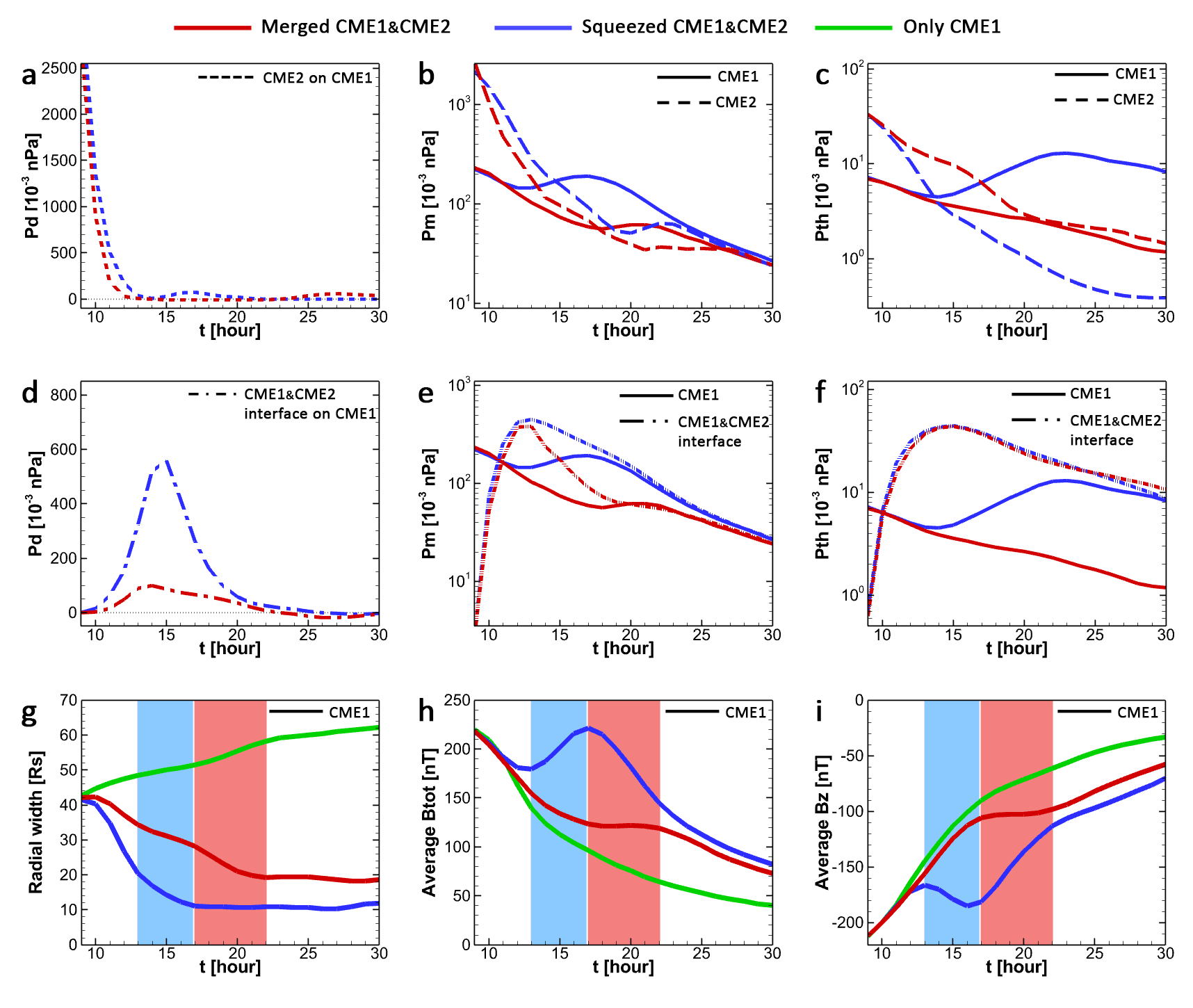}
    \caption{\textbf{Pressures acted on CME1 and the resulted magnetic field enhancement.} Panels a and d present the dynamic pressure ($P_d$) exerted on CME1 by CME2 and by the CME1-CME2 interface as functions of time. Panels b, c, e, and f show the time evolution of the magnetic pressure ($P_m$) and thermal pressure ($P_{\rm th}$) of CME1 (solid lines), CME2 (dashed lines), and the CME1-CME2 interface (dash-dotted lines). Panels g-i display the time variations of the radial width, the average total magnetic field magnitude ($B_{\rm tot}$), and the average magnetic field component $B_z$ of CME1 along the Sun-Earth line. Red lines correspond to the ``Merged CME1 and CME2'' simulation case, blue lines to the ``Squeezed CME1 and CME2'' case, and green lines to the ``Only CME1'' case. The shaded regions indicate the time interval during which the shock driven by CME2 crosses CME1 for the ``Merged CME1 and CME2'' (red region) and ``Squeezed CME1 and CME2''(blue region) cases.}\label{fig7}
\end{figure}

Fig.~\ref{fig7} shows that at different stages of the CME-CME interaction, different pressures act on the preceding CME, resulting in different degrees of variation in its properties. At the moment when the front of CME2 first contacts the rear of CME1 ($t=9$ h), the dynamic pressure of CME2 is the dominant force exerted on CME1 (Fig.~\ref{fig7}a), although both the magnetic pressure and thermal pressure of CME2 are much higher than those of CME1 (Fig.~\ref{fig7}b-c). As CME2 continuously chases CME1, CME2 decelerates and CME1 accelerates, which leads to a sharp decrease in the dynamic pressure of CME2 (Fig.~\ref{fig7}a). During this stage (from about $t=9$ h to $t=12$ h), the radial width of CME1 decreases rapidly (Fig.~\ref{fig7}g).

After $t=12$ h, the CME1-CME2 interface becomes the main driver of the variations in CME1 (Fig.~\ref{fig7}d-f), primarily through two stages.
In the first stage (from about $t=12$ h to $t=20$ h), the magnetic pressure of the interface is larger than that of CME1 (Fig.~\ref{fig7}e), 
and this magnetic pressure difference exceeds the thermal pressure difference between the interface and CME1 (Fig.~\ref{fig7}f), 
but is of the same order as the dynamic pressure of the interface (Fig.~\ref{fig7}d). 
Compared to the ``Only CME1'' case, this stage exhibits a rapid reduction in the radial width of CME1 and a clear enhancement of $B_{\rm tot}$ 
and the southward $B_z$ of CME1 for the two interaction cases (Fig.~\ref{fig7}g-i).

In the second stage (after about $t=20$ h), the magnetic pressure of the interface becomes nearly equal to that of CME1, and the thermal pressure of the interface becomes an important factor in maintaining the short radial extent of CME1 (Fig.~\ref{fig7}g) and slowing the decline of  
$B_z$ with propagation distance. For the ``Squeezed CME1 and CME2'' case, the thermal pressure of the interface becomes nearly equal to that of CME1 after about $t=25$ h, indicating that CME1 is no longer being actively squeezed, but retains its compressed state as it continues to propagate.
For the two interaction cases, $B_{\rm tot}$ and the southward 
$B_z$ of CME1 remain approximately twice those in the ``Only CME1'' case (Fig.~\ref{fig7}h-i), thereby generating sustained influences on Earth (Fig.~\ref{fig5}).
The comparison between the ``Merged CME1 and CME2'' case and the ``Squeezed CME1 and CME2'' case shows that 
the stronger dynamic pressure and magnetic pressure of CME2 produce a shorter radial width and a stronger $B_{\rm tot}$ and southward $B_z$ of CME1. 
 
\section{Summary and Discussion}
\label{sec:discuss}
Our combined observational and numerical analysis of the 10 May 2024 CME-CME interaction event reveals that mutual squeezing between successive CMEs results in the sustained enhancement of the southward magnetic field. Three homologous CMEs erupted from active region AR 13664 within a single day (Figs.~\ref{fig1}). Their successive propagation results in the trailing CME2 compressing the preceding CME1, reshaping the initially bulky cloud of CME1 into a flattened ``pancake'' morphology (Figs.~\ref{fig3} and ~\ref{fig4}). This squeezing not only prevents the natural expansion-induced decay of the magnetic field but also maintains a strong southward field (exceeding 60~nT at 1~AU) (Figs.~\ref{fig5} and \ref{fig6}), directly causing the extreme geomagnetic storm (Fig.~\ref{fig5}). 

The critical role of squeezing is further supported by the pressure analysis (Fig.~\ref{fig7}): after the initial dynamic-pressure-dominated contact, the CME1-CME2 interface exerts a persistent magnetic and thermal pressure that sustains the compressed state and the amplified southward field. Stronger squeezing leads to a stronger field enhancement (Fig.~\ref{fig7}). 
Because squeezing is an inevitable consequence of the catching-up process in successive CMEs, our identified mechanism reveals a natural link between CME-CME interaction and the generation of a strong southward magnetic field.
 
Our analysis, based on systematically comparing the ``Merged'' (same helicity) and ``Squeezed'' (opposite helicity) scenarios, demonstrates that squeezing is the dominant mechanism that both compresses the field and sustains it. The merging process moderates the squeezing effect to some extent. We do not exclude the possibility that other processes, such as interaction with the ambient solar wind \citep{Mayank2024, Lichen2026}, magnetic reconnection within CME flux ropes \citep{Gou2023, Schmieder2024}, or turbulence and wave activities inside CMEs \citep{Li2016, Li2017, Biswas2026}, also contribute to the amplification of the southward field. However, under similar conditions, the typical amplitudes of these factors are expected to be smaller than the squeezing-induced enhancement.

Several limitations of this study should be pointed out.  
First, our conclusions are derived from the observations and realistic simulations of the 10 May 2024 interaction event, and we have only varied the helicity of CME2 to isolate the effects of merging. Although the trailing CME inevitably catches up with the preceding one during interaction, how the initial separation, relative speed, propagation direction, and magnetic topology influence the squeezing process and the associated field enhancement remains unclear. 
Second, while our simulations reproduce most of the observed white-light and in-situ features, they yield an earlier arrival times of the CMEs at 1~AU, a common issue in prediction studies (e.g., \cite{Riley2018}). Dedicated adjustments to the background solar wind simulation or the CME initialization are needed to better match the observed arrival times. 
Third, the interaction among the CMEs results in a complex structure at 1 AU, which causes each CME to lose most of its individual features. Here, we combine the numerical results with the observations to recognize the likely intervals of CME1 and CME2, which show some differences from previous studies \citep{Weiler2025, Pal2025}. 

The CME mass can influence CME--CME interaction by affecting the tilt evolution under ambient magnetic torque \citep{Asvestari2022}, momentum transfer and collision characteristics \citep{Temmer2014}, and the drag from the ambient solar wind \citep{Werner2019,Yordanova2024}. 
We estimate the CME mass using the Thomson scattering approximation from white-light observations (e.g., \cite{Vourlidas2000, Temmer2021}):
\(M_{\rm obs} \approx m_p \sum_{i} [I_{{\rm pixel},i} / I_e(\theta_i)] \Delta A_i\),
where \(I_{\rm pixel}\) is the excess white-light brightness, \(I_e(\theta)\) is the theoretical Thomson scattering brightness per electron, and \(m_p\) is the proton mass. 
The observed CME masses are on the order of \(10^{12}\)--\(10^{13}\) kg.
In our simulation, integrating the density over the spheromak volume yields CME masses of approximately \( 2.25 \times 10^{13}\) kg, consistent with the observational estimates within a factor of \(\sim 2\). A dedicated follow-up study with explicit mass constraints would be valuable for further quantifying how the CME mass regulates flux-rope expansion and compression.

We inserted the CMEs into a steady-state background solar wind without a multi-day preconditioning run, in order to focus on the interaction among the three 8 May CMEs. An observational check of the preceding five days shows that most earlier activity was either not Earth-directed or consisted of slow CMEs (\(\sim\)311--353 km s\(^{-1}\)), which are expected to exert a weaker drag effect and provide limited preconditioning \citep{Yashiro2005}. Previous studies typically insert CMEs five days in advance to build up preconditioning \citep{Temmer2017, Pomoell2018}; incorporating all preceding events would further quantify its role, and we identify this as a limitation of the present study and a future direction.

Several additional diagnostics could further support the squeezing mechanism, including computing the CME aspect ratio to quantify the degree of squeezing, tracking the CME expansion and magnetic flux, analyzing the influence of reconnection between CMEs, and isolating the compressive effect of the shock on the CME. We also note that a systematic comparison of sheaths between squeezed CMEs and ordinary CME would be valuable and is left for future work.

Future work also focus on two complementary directions. First, a statistical survey of multiple CME interaction events is needed to establish the observational prevalence of squeezing-dominated southward field enhancement. This requires linking remote-sensing diagnostics of CME morphology from coronagraph and heliospheric imagers (e.g., from WISPR (Wide-field Imager for Parker Solar Probe), SoloHI (Solar Orbiter Heliospheric Imager), PUNCH (Polarimeter to Unify the Corona and Heliosphere)) to in-situ measurements (e.g., from Parker Solar Probe, Solar Orbiter, and near-Earth monitors) \citep{Hess2023}. Second, numerical simulations need to be further improved to capture realistic CME magnetic fields through a seamless modelling framework that tracks a CME from its birth in an active region through its subsequent propagation in the  solar corona and solar wind \citep{Jiang2026, Ye2026}.
Future multi-viewpoint observations from both the ecliptic plane and the polar region \citep{Deng2025} are expected to test whether the squeezing mechanism operates across different CME interaction geometries and heliospheric distances to produce southward magnetic field enhancement near Earth.



\begin{acknowledgments}
This work is supported by the National Natural Science Foundation of China (NSFC) (No. 42530206), the National Key R\&D Program
of China (No. 2022YFF0503800), China’s Space Origins Exploration Program, NSFC (Nos. 42274213, 42474216, 42530105,
42241118), and Project Supported by the Specialized Research Fund for State Key Laboratory of Solar Activity and Space Weather. R.W. also acknowledges support from the Strategic Priority Research Program of the Chinese Academy of Sciences (No. XDB0560000) and the Specialized Research Fund of CMA Key Laboratory of Space Weather. We acknowledge the solar wind measurements from WIND and remote sensing observations from STEREO and SDO.  We also thank the SUTRI team (NAOC \& PKU) for providing the EUV data. The numerical simulations were performed on the high-performance computing cluster at National Supercomputing Center in Tianjin (NSCC-TJ).   
\end{acknowledgments} 

\bibliography{ref}{}

\begin{thebibliography}{}
\expandafter\ifx\csname natexlab\endcsname\relax\def\natexlab#1{#1}\fi
\providecommand{\url}[1]{\href{#1}{#1}}
\providecommand{\dodoi}[1]{doi:~\href{http://doi.org/#1}{\nolinkurl{#1}}}
\providecommand{\doeprint}[1]{\href{http://ascl.net/#1}{\nolinkurl{http://ascl.net/#1}}}
\providecommand{\doarXiv}[1]{\href{https://arxiv.org/abs/#1}{\nolinkurl{https://arxiv.org/abs/#1}}}

\bibitem[{{Asvestari} {et~al.}(2022){Asvestari}, {Rindlisbacher}, {Pomoell}, \&
  {Kilpua}}]{Asvestari2022}
{Asvestari}, E., {Rindlisbacher}, T., {Pomoell}, J., \& {Kilpua}, E. K.~J.
  2022, \apj, 926, 87, \dodoi{10.3847/1538-4357/ac3a73}

\bibitem[{Behannon {et~al.}(1991)Behannon, Burlaga, \& Hewish}]{Behannon1991}
Behannon, K.~W., Burlaga, L.~F., \& Hewish, A. 1991, Journal of Geophysical
  Research: Space Physics, 96, 21213, \dodoi{10.1029/91ja02267}

\bibitem[{{Biswas} {et~al.}(2026){Biswas}, {Bhaskar}, {Abitha}, {Dhamane},
  {Pal}, {Chakrabarty}, \& {Yadav}}]{Biswas2026}
{Biswas}, S., {Bhaskar}, A., {Abitha}, S.~G., {et~al.} 2026, \apj, 1003, 230,
  \dodoi{10.3847/1538-4357/ae64e2}

\bibitem[{Biswas {et~al.}(2025)Biswas, Bhaskar, Raghav, Kumar, Ghag, Thampi, \&
  Yadav}]{Biswas2025}
Biswas, S., Bhaskar, A., Raghav, A., {et~al.} 2025, The Astrophysical Journal
  Letters, 991, L15, \dodoi{10.3847/2041-8213/adfe60}

\bibitem[{{Borovsky} \& {Denton}(2006)}]{Borovsky2006}
{Borovsky}, J.~E., \& {Denton}, M.~H. 2006, \jgr, 111, A07S08,
  \dodoi{10.1029/2005JA011447}

\bibitem[{{Burlaga} {et~al.}(1987){Burlaga}, {Behannon}, \&
  {Klein}}]{Burlaga1987}
{Burlaga}, L.~F., {Behannon}, K.~W., \& {Klein}, L.~W. 1987, \jgr, 92, 5725,
  \dodoi{10.1029/JA092iA06p05725}

\bibitem[{Chen {et~al.}(2024)Chen, Liu, Zhu, Hu, \& Wang}]{Chen2024}
Chen, C., Liu, Y.~D., Zhu, B., Hu, H., \& Wang, R. 2024, The Astrophysical
  Journal Letters, 969, L4, \dodoi{10.3847/2041-8213/ad53ca}

\bibitem[{Chen(2011)}]{Chen_2011}
Chen, P.~F. 2011, Living Reviews in Solar Physics, 8,
  \dodoi{10.12942/lrsp-2011-1}

\bibitem[{Chen {et~al.}(2025)Chen, Aa, Yuan, \& Zhang}]{Chen2025}
Chen, Y., Aa, E., Yuan, T., \& Zhang, S. 2025, National Science Review, 12,
  \dodoi{10.1093/nsr/nwaf307}

\bibitem[{Chi {et~al.}(2024)Chi, Shen, Zhang, Xu, Mao, Liu, Wang, Yu, Luo,
  Zhong, \& Wang}]{Chi2024}
Chi, Y., Shen, C., Zhang, Z., {et~al.} 2024, \apjl, 975, L25,
  \dodoi{10.3847/2041-8213/ad87e8}

\bibitem[{{Dai} {et~al.}(2024){Dai}, {Zhu}, {Ren}, {Gonzalez}, {Wang},
  {Sibeck}, {Samsonov}, {Escoubet}, {Tang}, {Zhang}, \&
  {Branduardi-Raymont}}]{Dai2024}
{Dai}, L., {Zhu}, M., {Ren}, Y., {et~al.} 2024, Nature Communications, 15, 639,
  \dodoi{10.1038/s41467-024-44992-y}

\bibitem[{Dasso {et~al.}(2009)Dasso, Mandrini, Schmieder, Cremades, Cid,
  Cerrato, Saiz, Démoulin, Zhukov, Rodriguez, Aran, Menvielle, \&
  Poedts}]{Dasso2009}
Dasso, S., Mandrini, C.~H., Schmieder, B., {et~al.} 2009, Journal of
  Geophysical Research: Space Physics, 114, \dodoi{10.1029/2008ja013102}

\bibitem[{Deng {et~al.}(2025)Deng, Tian, Jiang, Yang, Li, Cameron, Gizon,
  Harra, {Wimmer-Schweingruber}, Auch{\`e}re, Bai, Bellot, Chen, Chen, Chitta,
  Davies, Favata, Feng, Feng, Gan, Hassler, He, Hou, Hou, Jin, Li, Lin, Nandy,
  Pant, Romoli, Sakao, Krishna~Prasad, Shen, Su, Toriumi, Tripathi, Wang, Wang,
  Xia, Xiong, Yan, Yang, Yang, Zhang, Zhou, Zhu, Wang, \& Wang}]{Deng2025}
Deng, Y., Tian, H., Jiang, J., {et~al.} 2025, Chinese Journal of Space Science,
  45, 913, \dodoi{10.11728/cjss2025.04.2025-0054}

\bibitem[{{Farrugia} \& {Berdichevsky}(2004)}]{Farrugia2004}
{Farrugia}, C., \& {Berdichevsky}, D. 2004, Annales Geophysicae, 22, 3679,
  \dodoi{10.5194/angeo-22-3679-2004}

\bibitem[{Feng {et~al.}(2012)Feng, Yang, Xiang, Jiang, Ma, Wu, Zhong, \&
  Zhou}]{Feng2012}
Feng, X., Yang, L., Xiang, C., {et~al.} 2012, Solar Physics, 279, 207,
  \dodoi{10.1007/s11207-012-9969-9}

\bibitem[{{Fu} {et~al.}(2025){Fu}, {Fu}, {Zhang}, {Yu}, \& {Cao}}]{Fu2025}
{Fu}, W.~D., {Fu}, H.~S., {Zhang}, W.~Z., {Yu}, Y., \& {Cao}, J.~B. 2025, \grl,
  52, 2024GL114040, \dodoi{10.1029/2024GL114040}

\bibitem[{{Gopalswamy} {et~al.}(2025){Gopalswamy}, {Akiyama}, {Yashiro},
  {M{\"a}kel{\"a}}, \& {Xie}}]{Gopalswamy2025}
{Gopalswamy}, N., {Akiyama}, S., {Yashiro}, S., {M{\"a}kel{\"a}}, P., \& {Xie},
  H. 2025, Journal of Astrophysics and Astronomy, 46, 78,
  \dodoi{10.1007/s12036-025-10091-7}

\bibitem[{{Gopalswamy} {et~al.}(2001){Gopalswamy}, {Yashiro}, {Kaiser},
  {Howard}, \& {Bougeret}}]{Gopalswamy2001}
{Gopalswamy}, N., {Yashiro}, S., {Kaiser}, M.~L., {Howard}, R.~A., \&
  {Bougeret}, J.-L. 2001, \apjl, 548, L91, \dodoi{10.1086/318939}

\bibitem[{Gou {et~al.}(2023)Gou, Liu, Veronig, Zhuang, Li, Wang, Xu, \&
  Wang}]{Gou2023}
Gou, T., Liu, R., Veronig, A.~M., {et~al.} 2023, Nature Astronomy, 7, 815,
  \dodoi{10.1038/s41550-023-01966-2}

\bibitem[{Hayakawa {et~al.}(2025)Hayakawa, Ebihara, Mishev, Koldobskiy, Kusano,
  Bechet, Yashiro, Iwai, Shinbori, Mursula, Miyake, Shiota, Silveira, Stuart,
  Oliveira, Akiyama, Ohnishi, Ledvina, \& Miyoshi}]{Hayakawa2025}
Hayakawa, H., Ebihara, Y., Mishev, A., {et~al.} 2025, The Astrophysical
  Journal, 979, 49, \dodoi{10.3847/1538-4357/ad9335}

\bibitem[{Hess {et~al.}(2023)Hess, Colaninno, Vourlidas, Howard, \&
  Stenborg}]{Hess2023}
Hess, P., Colaninno, R.~C., Vourlidas, A., Howard, R.~A., \& Stenborg, G. 2023,
  Astronomy \& Astrophysics, 679, A149, \dodoi{10.1051/0004-6361/202346907}

\bibitem[{{Ivanov}(1982)}]{Ivanov1982}
{Ivanov}, K.~G. 1982, \ssr, 32, 49, \dodoi{10.1007/BF00225176}

\bibitem[{{Jarolim} {et~al.}(2024){Jarolim}, {Veronig}, {Purkhart}, {Zhang}, \&
  {Rempel}}]{Jarolim2024}
{Jarolim}, R., {Veronig}, A.~M., {Purkhart}, S., {Zhang}, P., \& {Rempel}, M.
  2024, The Astrophysical Journal Letters, 976, L12,
  \dodoi{10.3847/2041-8213/ad8914}

\bibitem[{{Jiang} {et~al.}(2026){Jiang}, {Feng}, {Yang}, {Li}, {Guo}, {Zuo}, \&
  {Wang}}]{Jiang2026}
{Jiang}, C., {Feng}, X., {Yang}, L., {et~al.} 2026, \apj, 1004, 189,
  \dodoi{10.3847/1538-4357/ae69de}

\bibitem[{Khuntia {et~al.}(2025)Khuntia, Mishra, \& Agarwal}]{Khuntia2025}
Khuntia, S., Mishra, W., \& Agarwal, A. 2025, Astronomy \& Astrophysics, 698,
  A79, \dodoi{10.1051/0004-6361/202452866}

\bibitem[{Koehn {et~al.}(2022)Koehn, Desai, Davies, Forsyth, Eastwood, \&
  Poedts}]{Koehn2022}
Koehn, G.~J., Desai, R.~T., Davies, E.~E., {et~al.} 2022, The Astrophysical
  Journal, 941, 139, \dodoi{10.3847/1538-4357/aca28c}

\bibitem[{Kwak {et~al.}(2024)Kwak, Kim, Kim, Miyashita, Yang, Park, Lim, Jung,
  Kam, Lee, Lee, Yoo, Lee, Kwon, Seough, Nam, Lee, Hong, Sohn, \&
  Talha}]{Kwak2024}
Kwak, Y.~S., Kim, J.~H., Kim, S., {et~al.} 2024, Journal of Astronomy and Space
  Sciences, 41, 171, \dodoi{10.5140/jass.2024.41.3.171}

\bibitem[{{Li} {et~al.}(2016){Li}, {Wang}, {He}, {Zhang}, {Richardson},
  {Belcher}, \& {Tu}}]{Li2016}
{Li}, H., {Wang}, C., {He}, J., {et~al.} 2016, The Astrophysical Journal
  Letters, 831, L13, \dodoi{10.3847/2041-8205/831/2/L13}

\bibitem[{{Li} {et~al.}(2017){Li}, {Wang}, {Richardson}, \& {Tu}}]{Li2017}
{Li}, H., {Wang}, C., {Richardson}, J.~D., \& {Tu}, C. 2017, The Astrophysical
  Journal Letters, 851, L2, \dodoi{10.3847/2041-8213/aa9c3f}

\bibitem[{Li {et~al.}(2024)Li, Liu, Jing, Li, \& Chen}]{Li2024}
Li, Y., Liu, X., Jing, Z., Li, Y., \& Chen, W. 2024, The Astrophysical Journal
  Letters, 972, L1, \dodoi{10.3847/2041-8213/ad6d6c}

\bibitem[{{Li} {et~al.}(2026){Li}, {Chen}, {Guo}, {Guo}, {Wu}, {Huang},
  {Cheng}, {Ding}, \& {Keppens}}]{Lichen2026}
{Li}, Y., {Chen}, G., {Guo}, J., {et~al.} 2026, \apj, 997, 228,
  \dodoi{10.3847/1538-4357/ae27d1}

\bibitem[{{Liu} {et~al.}(2024){Liu}, {Hu}, {Zhao}, {Chen}, \& {Wang}}]{Liu2024}
{Liu}, Y.~D., {Hu}, H., {Zhao}, X., {Chen}, C., \& {Wang}, R. 2024, \apjl, 974,
  L8, \dodoi{10.3847/2041-8213/ad7ba4}

\bibitem[{{Liu} {et~al.}(2012){Liu}, {Luhmann}, {M{\"o}stl},
  {Martinez-Oliveros}, {Bale}, {Lin}, {Harrison}, {Temmer}, {Webb}, \&
  {Odstrcil}}]{Liu2012}
{Liu}, Y.~D., {Luhmann}, J.~G., {M{\"o}stl}, C., {et~al.} 2012, \apjl, 746,
  L15, \dodoi{10.1088/2041-8205/746/2/L15}

\bibitem[{{Liu} {et~al.}(2014){Liu}, {Luhmann}, {Kajdi{\v{c}}}, {Kilpua},
  {Lugaz}, {Nitta}, {M{\"o}stl}, {Lavraud}, {Bale}, {Farrugia}, \&
  {Galvin}}]{Liu2014}
{Liu}, Y.~D., {Luhmann}, J.~G., {Kajdi{\v{c}}}, P., {et~al.} 2014, Nature
  Communications, 5, 3481, \dodoi{10.1038/ncomms4481}

\bibitem[{{Lugaz} \& {Farrugia}(2014)}]{Lugaz2014}
{Lugaz}, N., \& {Farrugia}, C.~J. 2014, \grl, 41, 769,
  \dodoi{10.1002/2013GL058789}

\bibitem[{{Lugaz} {et~al.}(2015{\natexlab{a}}){Lugaz}, {Farrugia}, {Huang}, \&
  {Spence}}]{Lugaz2015}
{Lugaz}, N., {Farrugia}, C.~J., {Huang}, C.-L., \& {Spence}, H.~E.
  2015{\natexlab{a}}, \grl, 42, 4694, \dodoi{10.1002/2015GL064530}

\bibitem[{{Lugaz} {et~al.}(2015{\natexlab{b}}){Lugaz}, {Farrugia}, {Smith}, \&
  {Paulson}}]{Lugaz2015ab}
{Lugaz}, N., {Farrugia}, C.~J., {Smith}, C.~W., \& {Paulson}, K.
  2015{\natexlab{b}}, Journal of Geophysical Research (Space Physics), 120,
  2409, \dodoi{10.1002/2014JA020848}

\bibitem[{Lugaz {et~al.}(2007)Lugaz, Manchester~IV, Roussev, Toth, \&
  Gombosi}]{Lugaz2007}
Lugaz, N., Manchester~IV, W.~B., Roussev, I.~I., Toth, G., \& Gombosi, T.~I.
  2007, The Astrophysical Journal, 659, 788, \dodoi{10.1086/512005}

\bibitem[{Lugaz {et~al.}(2017)Lugaz, Temmer, Wang, \& Farrugia}]{Lugaz2017}
Lugaz, N., Temmer, M., Wang, Y., \& Farrugia, C.~J. 2017, Solar Physics, 292,
  \dodoi{10.1007/s11207-017-1091-6}

\bibitem[{{Luhmann} {et~al.}(2020){Luhmann}, {Gopalswamy}, {Jian}, \&
  {Lugaz}}]{Luhmann2020}
{Luhmann}, J.~G., {Gopalswamy}, N., {Jian}, L.~K., \& {Lugaz}, N. 2020,
  \solphys, 295, 61, \dodoi{10.1007/s11207-020-01624-0}

\bibitem[{Ma {et~al.}(2024)Ma, Yang, Shen, Shen, Chi, Wang, Zhou, Zhang,
  Heyner, Auster, Richter, \& {Sanchez-Cano}}]{Ma2024}
Ma, M., Yang, L., Shen, F., {et~al.} 2024, \apj, 976, 183,
  \dodoi{10.3847/1538-4357/ad8a5a}

\bibitem[{{Manchester} {et~al.}(2017){Manchester}, {Kilpua}, {Liu}, {Lugaz},
  {Riley}, {T{\"o}r{\"o}k}, \& {Vr{\v{s}}nak}}]{Manchester2017}
{Manchester}, W., {Kilpua}, E. K.~J., {Liu}, Y.~D., {et~al.} 2017, \ssr, 212,
  1159, \dodoi{10.1007/s11214-017-0394-0}

\bibitem[{Mayank {et~al.}(2024)Mayank, Lotz, Vaidya, Mishra, \&
  Chakrabarty}]{Mayank2024}
Mayank, P., Lotz, S., Vaidya, B., Mishra, W., \& Chakrabarty, D. 2024, The
  Astrophysical Journal, 976, 126, \dodoi{10.3847/1538-4357/ad8084}

\bibitem[{{Pal} {et~al.}(2025){Pal}, {Mac Cormack}, {Kilpua}, {Yogesh}, {Jian},
  \& {Nieves-Chinchilla}}]{Pal2025}
{Pal}, S., {Mac Cormack}, C., {Kilpua}, E. K.~J., {et~al.} 2025, \aap, 702,
  A150, \dodoi{10.1051/0004-6361/202555908}

\bibitem[{{Pomoell} \& {Poedts}(2018)}]{Pomoell2018}
{Pomoell}, J., \& {Poedts}, S. 2018, Journal of Space Weather and Space
  Climate, 8, A35, \dodoi{10.1051/swsc/2018020}

\bibitem[{{Razquin} {et~al.}(2026){Razquin}, {Dissauer}, {Veronig}, \&
  {Barnes}}]{Razquin2026}
{Razquin}, A., {Dissauer}, K., {Veronig}, A.~M., \& {Barnes}, G. 2026,
  Astronomy \& Astrophysics, 708, A292, \dodoi{10.1051/0004-6361/202659130}

\bibitem[{Ren {et~al.}(2026)Ren, Zhu, Dai, Gonzalez, Wang, \& Wang}]{Ren2026}
Ren, Y., Zhu, M., Dai, L., {et~al.} 2026, Geophysical Research Letters, 53,
  e2025GL119629, \dodoi{10.1029/2025GL119629}

\bibitem[{{Riley} {et~al.}(2018){Riley}, {Mays}, {Andries}, {Amerstorfer},
  {Biesecker}, {Delouille}, {Dumbovi{\'c}}, {Feng}, {Henley}, {Linker},
  {M{\"o}stl}, {Nu{\~n}ez}, {Pizzo}, {Temmer}, {Tobiska}, {Verbeke}, {West}, \&
  {Zhao}}]{Riley2018}
{Riley}, P., {Mays}, M.~L., {Andries}, J., {et~al.} 2018, Space Weather, 16,
  1245, \dodoi{10.1029/2018SW001962}

\bibitem[{{Sabri} {et~al.}(2025){Sabri}, {Mahmoudian}, \& {Poedts}}]{Sabri2025}
{Sabri}, S., {Mahmoudian}, A., \& {Poedts}, S. 2025, Journal of Geophysical
  Research (Space Physics), 130, e2024JA033554, \dodoi{10.1029/2024JA033554}

\bibitem[{Schmieder {et~al.}(2024)Schmieder, Guo, \& Poedts}]{Schmieder2024}
Schmieder, B., Guo, J., \& Poedts, S. 2024, Reviews of Modern Plasma Physics,
  8, 27, \dodoi{10.1007/s41614-024-00166-3}

\bibitem[{Scolini {et~al.}(2020)Scolini, Chan, Temmer, Kilpua, Dissauer,
  Veronig, Palmerio, Pomoell, Dumbovi, Guo, Rodriguez, \& Poedts}]{Scolini2020}
Scolini, C., Chan, E., Temmer, M., {et~al.} 2020, The Astrophysical Journal
  Supplement Series, 247, 21, \dodoi{10.3847/1538-4365/ab6216}

\bibitem[{{Shen} {et~al.}(2012){Shen}, {Wang}, {Wang}, {Liu}, {Liu},
  {Vourlidas}, {Miao}, {Ye}, {Liu}, \& {Zhou}}]{Shen2012}
{Shen}, C., {Wang}, Y., {Wang}, S., {et~al.} 2012, Nature Physics, 8, 923,
  \dodoi{10.1038/nphys2440}

\bibitem[{Song {et~al.}(2026)Song, Wang, Li, Shen, Wang, He, Xia, \&
  Chen}]{Song2026}
Song, H., Wang, R., Li, L., {et~al.} 2026, The Astrophysical Journal, 997, 344,
  \dodoi{10.3847/1538-4357/ae360d}

\bibitem[{{Temmer} {et~al.}(2017){Temmer}, {Reiss}, {Nikolic}, {Hofmeister}, \&
  {Veronig}}]{Temmer2017}
{Temmer}, M., {Reiss}, M.~A., {Nikolic}, L., {Hofmeister}, S.~J., \& {Veronig},
  A.~M. 2017, \apj, 835, 141, \dodoi{10.3847/1538-4357/835/2/141}

\bibitem[{{Temmer} {et~al.}(2014){Temmer}, {Veronig}, {Peinhart}, \&
  {Vr{\v{s}}nak}}]{Temmer2014}
{Temmer}, M., {Veronig}, A.~M., {Peinhart}, V., \& {Vr{\v{s}}nak}, B. 2014,
  \apj, 785, 85, \dodoi{10.1088/0004-637X/785/2/85}

\bibitem[{Temmer {et~al.}(2012)Temmer, Vršnak, Rollett, Bein, de~Koning, Liu,
  Bosman, Davies, Möstl, Žic, Veronig, Bothmer, Harrison, Nitta, Bisi, Flor,
  Eastwood, Odstrcil, \& Forsyth}]{Temmer2012}
Temmer, M., Vršnak, B., Rollett, T., {et~al.} 2012, The Astrophysical Journal,
  749, 57, \dodoi{10.1088/0004-637x/749/1/57}

\bibitem[{{Temmer} {et~al.}(2021){Temmer}, {Holzknecht}, {Dumbovi{\'c}},
  {Vr{\v{s}}nak}, {Sachdeva}, {Heinemann}, {Dissauer}, {Scolini}, {Asvestari},
  {Veronig}, \& {Hofmeister}}]{Temmer2021}
{Temmer}, M., {Holzknecht}, L., {Dumbovi{\'c}}, M., {et~al.} 2021, Journal of
  Geophysical Research: Space Physics, 126, \dodoi{10.1029/2020JA028380}

\bibitem[{{Thernisien} {et~al.}(2006){Thernisien}, {Howard}, \&
  {Vourlidas}}]{Thernisien2006}
{Thernisien}, A.~F.~R., {Howard}, R.~A., \& {Vourlidas}, A. 2006, \apj, 652,
  763, \dodoi{10.1086/508254}

\bibitem[{Tsurutani {et~al.}(1988)Tsurutani, Gonzalez, Tang, Akasofu, \&
  Smith}]{Tsurutani1988}
Tsurutani, B.~T., Gonzalez, W.~D., Tang, F., Akasofu, S.~I., \& Smith, E.~J.
  1988, Journal of Geophysical Research: Space Physics, 93, 8519,
  \dodoi{10.1029/ja093ia08p08519}

\bibitem[{{Vourlidas} {et~al.}(2019){Vourlidas}, {Patsourakos}, \&
  {Savani}}]{Vourlidas2019}
{Vourlidas}, A., {Patsourakos}, S., \& {Savani}, N.~P. 2019, Philosophical
  Transactions of the Royal Society of London Series A, 377, 20180096,
  \dodoi{10.1098/rsta.2018.0096}

\bibitem[{{Vourlidas} {et~al.}(2000){Vourlidas}, {Subramanian}, {Dere}, \&
  {Howard}}]{Vourlidas2000}
{Vourlidas}, A., {Subramanian}, P., {Dere}, K.~P., \& {Howard}, R.~A. 2000,
  \apj, 534, 456, \dodoi{10.1086/308747}

\bibitem[{Wang {et~al.}(2024)Wang, Liu, Zhao, \& Hu}]{WangR2024}
Wang, R., Liu, Y.~D., Zhao, X., \& Hu, H. 2024, Astronomy \& Astrophysics, 692,
  A112, \dodoi{10.1051/0004-6361/202452008}

\bibitem[{{Wang} {et~al.}(2003){Wang}, {Ye}, \& {Wang}}]{Wang2003a}
{Wang}, Y.~M., {Ye}, P.~Z., \& {Wang}, S. 2003, Journal of Geophysical Research
  (Space Physics), 108, 1370, \dodoi{10.1029/2003JA009850}

\bibitem[{Weiler {et~al.}(2025)Weiler, Möstl, Davies, Veronig, Amerstorfer,
  Amerstorfer, Le~Louëdec, Bauer, Lugaz, Haberle, Rüdisser, Majumdar, \&
  Reiss}]{Weiler2025}
Weiler, E., Möstl, C., Davies, E.~E., {et~al.} 2025, Space Weather, 23,
  \dodoi{10.1029/2024sw004260}

\bibitem[{{Werner} {et~al.}(2019){Werner}, {Yordanova}, {Dimmock}, \&
  {Temmer}}]{Werner2019}
{Werner}, A.~L.~E., {Yordanova}, E., {Dimmock}, A.~P., \& {Temmer}, M. 2019,
  Space Weather, 17, 357, \dodoi{10.1029/2018SW001993}

\bibitem[{Wijsen {et~al.}(2025)Wijsen, Jebaraj, Dresing, Kouloumvakos,
  Palmerio, \& Rodríguez-García}]{Wijsen2025}
Wijsen, N., Jebaraj, I.~C., Dresing, N., {et~al.} 2025, Astronomy \&
  Astrophysics, 699, A51, \dodoi{10.1051/0004-6361/202453598}

\bibitem[{Xiong {et~al.}(2009)Xiong, Zheng, \& Wang}]{xiong2009}
Xiong, M., Zheng, H., \& Wang, S. 2009, Journal of Geophysical Research: Space
  Physics, 114, \dodoi{10.1029/2009ja0140799}

\bibitem[{Xiong {et~al.}(2007)Xiong, Zheng, Wu, Wang, \& Wang}]{Xiong2007}
Xiong, M., Zheng, H., Wu, S.~T., Wang, Y., \& Wang, S. 2007, Journal of
  Geophysical Research: Space Physics, 112, A11, \dodoi{10.1029/2007ja012320}

\bibitem[{Xu {et~al.}(2019)Xu, Shen, Wang, Luo, \& Chi}]{Xu2019}
Xu, M., Shen, C., Wang, Y., Luo, B., \& Chi, Y. 2019, The Astrophysical Journal
  Letters, 884, L30, \dodoi{10.3847/2041-8213/ab4717}

\bibitem[{{Xue} {et~al.}(2005){Xue}, {Wang}, {Ye}, {Wang}, \&
  {Xiong}}]{Xue2005}
{Xue}, X.~H., {Wang}, Y., {Ye}, P.~Z., {Wang}, S., \& {Xiong}, M. 2005,
  Planetary and Space Science, 53, 443, \dodoi{10.1016/j.pss.2004.10.002}

\bibitem[{Yang {et~al.}(2012)Yang, Feng, Xiang, Liu, Zhao, \& Wu}]{Yang2012}
Yang, L.~P., Feng, X.~S., Xiang, C.~Q., {et~al.} 2012, Journal of Geophysical
  Research: Space Physics, 117, A8, \dodoi{10.1029/2011ja017494}

\bibitem[{Yang {et~al.}(2021)Yang, Wang, Feng, Xiong, Zhang, Zhu, Li, Zhou,
  Shen, Zhao, \& Liu}]{Yang2021}
Yang, L.~P., Wang, H., Feng, X., {et~al.} 2021, The Astrophysical Journal, 918,
  31, \dodoi{10.3847/1538-4357/ac0ef7}

\bibitem[{Yang {et~al.}(2023)Yang, Hou, Feng, He, Xiong, Zhang, Zhou, Shen,
  Zhao, Li, Yang, \& Liu}]{Yang2023}
Yang, L.~P., Hou, C., Feng, X.~S., {et~al.} 2023, The Astrophysical Journal,
  942, 65, \dodoi{10.3847/1538-4357/aca52d}

\bibitem[{Yang {et~al.}(2025)Yang, Feng, Shen, Xiong, Shen, Chi, Wang, Yan, Ma,
  Zhou, Zhang, \& Zhao}]{Yang2025}
Yang, L.~P., Feng, X.~S., Shen, F., {et~al.} 2025, The Astrophysical Journal,
  981, 109, \dodoi{10.3847/1538-4357/adb313}

\bibitem[{{Yashiro} {et~al.}(2005){Yashiro}, {Gopalswamy}, {Akiyama},
  {Michalek}, \& {Howard}}]{Yashiro2005}
{Yashiro}, S., {Gopalswamy}, N., {Akiyama}, S., {Michalek}, G., \& {Howard},
  R.~A. 2005, Journal of Geophysical Research: Space Physics, 110,
  \dodoi{10.1029/2005JA011151}

\bibitem[{{Ye} {et~al.}(2026){Ye}, {Shen}, {Raymond}, {Cai}, {Mei}, {Hu},
  {Chen}, {Li}, \& {Lin}}]{Ye2026}
{Ye}, J., {Shen}, C., {Raymond}, J.~C., {et~al.} 2026, Nature Communications,
  17, 8150, \dodoi{10.1038/s41467-026-75039-z}

\bibitem[{{Yordanova} {et~al.}(2024){Yordanova}, {Temmer}, {Dumbovi{\'c}},
  {Scolini}, {Paouris}, {Werner}, {Dimmock}, \&
  {Sorriso-Valvo}}]{Yordanova2024}
{Yordanova}, E., {Temmer}, M., {Dumbovi{\'c}}, M., {et~al.} 2024, Space
  Weather, 22, \dodoi{10.1029/2023SW003497}

\bibitem[{{Zhang} {et~al.}(2021){Zhang}, {Temmer}, {Gopalswamy}, {Malandraki},
  {Nitta}, {Patsourakos}, {Shen}, {Vr{\v{s}}nak}, {Wang}, {Webb}, {Desai},
  {Dissauer}, {Dresing}, {Dumbovi{\'c}}, {Feng}, {Heinemann}, {Laurenza},
  {Lugaz}, \& {Zhuang}}]{Zhang2021}
{Zhang}, J., {Temmer}, M., {Gopalswamy}, N., {et~al.} 2021, Progress in Earth
  and Planetary Science, 8, 56, \dodoi{10.1186/s40645-021-00426-7}

\bibitem[{{Zhang} {et~al.}(2026{\natexlab{a}}){Zhang}, {Liu}, {Li}, {Du},
  {Wang}, {Chen}, \& {Le}}]{Zhang2026ab}
{Zhang}, R., {Liu}, L., {Li}, W., {et~al.} 2026{\natexlab{a}}, \grl, 53,
  e2025GL118472, \dodoi{10.1029/2025GL118472}

\bibitem[{{Zhang} {et~al.}(2026{\natexlab{b}}){Zhang}, {Yang}, {Feng}, {Tian},
  {Ma}, {Shen}, {He}, {Zhang}, {Zhou}, {Wang}, {Ma}, \& {Zhang}}]{Zhang2026}
{Zhang}, X., {Yang}, L., {Feng}, X., {et~al.} 2026{\natexlab{b}}, \apj, 1000,
  200, \dodoi{10.3847/1538-4357/ae4c54}

\end{thebibliography}
\bibliographystyle{aasjournal}



\end{document}